\documentclass[aps,prb,twocolumn,superscriptaddress,footinbib]{revtex4}
\usepackage{graphicx}
\usepackage{units}
\newcommand{\footnoteremember}[2]{
\footnote{#2}
\newcounter{#1}
\setcounter{#1}{\value{footnote}}
}
\newcommand{\footnoterecall}[1]{
\footnotemark[\value{#1}]
}
\usepackage[latin1]{inputenc}

\newcommand{\subscript}[1]{\ensuremath{_{\textrm{#1}}}}

\begin{document}
\title{Localized state and charge transfer in nitrogen-doped graphene}
\author{Frédéric Joucken}
\email[Corresponding author: ]{fjoucken@fundp.ac.be}
\affiliation{Research Center in Physics of Matter and Radiation (PMR), University of Namur (FUNDP), 61 Rue de Bruxelles, 5000 Namur, Belgium}
\author{Yann Tison}
\author{J\'{e}r\^{o}me Lagoute}
\affiliation{Laboratoire Mat\'{e}riaux et Ph\'{e}nom\`{e}nes Quantiques, Universit\'{e} Paris Diderot Paris 7, Sorbonne Paris Cit\'{e}, CNRS, UMR 7162 case courrier 7021, 75205 Paris 13, France}
\author{Jacques Dumont}
\author{Damien Cabosart}
\altaffiliation{Present address: Universit\'e catholique de Louvain (UCL), Institute of Condensed Matter and Nanosciences (IMCN), 1/6 Place L. Pasteur, 1348 Louvain-la-Neuve, Belgium}
\affiliation{Research Center in Physics of Matter and Radiation (PMR), University of Namur (FUNDP), 61 Rue de Bruxelles, 5000 Namur, Belgium}
\author{Bing Zheng}
\altaffiliation[Present address: ] {Department Materials Science and Engineering, University of Wisconsin-Madison, Madison, WI 53706-1595, USA  }
\affiliation{Universit\'e catholique de Louvain (UCL), Institute of Condensed Matter and Nanosciences (IMCN), 1/6 Place L. Pasteur, 1348 Louvain-la-Neuve, Belgium}
\author{Vincent Repain}
\author{Cyril Chacon}
\author{Yann Girard}
\affiliation{Laboratoire Mat\'{e}riaux et Ph\'{e}nom\`{e}nes Quantiques, UMR7162, Universit\'{e} Paris Diderot Paris 7, Sorbonne Paris Cit\'{e}, CNRS, UMR 7162 case courrier 7021, 75205 Paris 13, France}
\author{Andr\'{e}s Rafael Botello-M\'{e}ndez}
\affiliation{Universit\'e catholique de Louvain (UCL), Institute of Condensed Matter and Nanosciences (IMCN), 1/6 Place L. Pasteur, 1348 Louvain-la-Neuve, Belgium}
\author{Sylvie Rousset}
\affiliation{Laboratoire Mat\'{e}riaux et Ph\'{e}nom\`{e}nes Quantiques, UMR7162, Universit\'{e} Paris Diderot Paris 7, Sorbonne Paris Cit\'{e}, CNRS, UMR 7162 case courrier 7021, 75205 Paris 13, France}
\author{Robert Sporken}
\affiliation{Research Center in Physics of Matter and Radiation (PMR), University of Namur (FUNDP), 61 Rue de Bruxelles, 5000 Namur, Belgium}
\author{Jean-Christophe Charlier}
\affiliation{Universit\'e catholique de Louvain (UCL), Institute of Condensed Matter and Nanosciences (IMCN), 1/6 Place L. Pasteur, 1348 Louvain-la-Neuve, Belgium}
\author{Luc Henrard}
\affiliation{Research Center in Physics of Matter and Radiation (PMR), University of Namur (FUNDP), 61 Rue de Bruxelles, 5000 Namur, Belgium}

 \begin{abstract}
  Nitrogen-doped epitaxial graphene grown
  on SiC(000\={1}) was prepared by exposing the surface to an atomic
  nitrogen flux. Using Scanning Tunneling Microscopy (STM) and
  Spectroscopy (STS), supported by Density Functional Theory (DFT) calculations, the simple substitution
    of carbon by nitrogen atoms has been identified as the most common
    doping configuration. High-resolution images reveal a
  reduction of local charge density on top of the nitrogen atoms,
  indicating a charge transfer to the neighboring carbon atoms. For
  the first time, local STS spectra clearly evidenced the
  energy levels associated with the chemical doping by nitrogen,
  localized in the conduction band. Various other nitrogen-related
  defects have been observed. The bias dependence of their
  topographic signatures  demonstrates the presence of
  structural configurations more complex than substitution as well as
  hole-doping.
\end{abstract}

\maketitle

Graphene has been proposed as a promising alternative to silicon-based
electronics for some applications. However a reliable control of its electronic properties,
for example by chemical doping, is still a challenging
task~\cite{neto,liu,fuhrer}. For carbon-based
materials, the incorporation of nitrogen in the lattice is a natural
choice because of its ability to form covalent bonds and
to modify the electronic properties of sp$^2$ carbon locally, with
minor structural perturbations~\cite{ewels,lherbier,zheng}. N-doped graphene
also offers interesting prospects for various other applications
including biosensing~\cite{wang}, field emission~\cite{soin}, lithium
incorporation~\cite{reddy,cho} or transparent electrodes~\cite{kasry}.

The substitution of some carbon atoms by nitrogen is expected to give
rise to donor states and then to n-type doping~\cite{zheng,wei}. The synthesis of chemically modified graphene
has been achieved either by direct growth of modified
layers~\cite{zhao,panchakarla,wei,meyer} or by post-growth treatment
of pristine graphene~\cite{guo,lin,jeong,wang,soin}. These studies
revealed the presence of several atomic configurations for the nitrogen
atoms: substitutional (`graphitic'), pyridine-like, or
pyrrole-like N~\cite{soin}.

However, a clear correlation between the synthesis methods and the atomic
configuration of the chemically modified graphene, on the one hand, and
between the atomic configuration and the electronic properties, on the
other hand, remains a challenging task. A step in this direction has
been achieved very recently by Zhao \textit{et al.}~\cite{zhao} through
Scanning Tunneling Microscopy (STM) and Spectroscopy (STS)
investigations of nitrogen-doped graphene prepared by chemical vapor
deposition on a copper substrate with NH$_3$ gas in the
feedstock. These authors determined the atomic configuration of the
nitrogen atoms to be predominantly (90\%) a simple substitution
(``graphitic'' nitrogen), with the majority of dopants located on the
same carbon sublattice of graphene. This work left open some key
questions relative to the interpretation of the experimental
data. For instance, simulated STM images exhibit a depletion above the
N atom (see also Zheng \textit{et al.}~\cite{zheng}) whereas experimental results have no such central feature. More importantly, the N-induced donor
energy level has not been evidenced in their STS measurements.

We present here a STM/STS study of N-doped graphene samples on
SiC(000\={1}) obtained by post-synthesis treatment. Our STM
images provide clear evidence of the presence of substitutional
nitrogen atoms together with more complex structures presenting well-defined
topographical features. Local spectroscopy, supported by simulations based on Density Functional Theory (DFT),
reveals for the first time a localized donor state in
the electronic density of states related to the substitutional
nitrogen atoms. Based on STM images at different biases, we also observed and analysed a change in the STM topographic image with respect to the tip-graphene distance that reconciles experiment and simulation. Doping configurations different from substitution are also analyzed.

Our graphene samples were prepared from C-terminated n-type
6H-SiC(000\={1}) wafers, following procedures available in the
literature~\cite{forbeaux,deheer,varchon,hiebel,starke} which lead to
multilayered graphene with misoriented, decoupled layers\footnote{See Supplemental Material at [...] and references 1-9 therein for details on the sample preparation and on the sample properties.}. N-doping was achieved by exposure to an atomic nitrogen flux produced by a remote radiofrequency plasma source fed with N$_2$. The samples were then analyzed using a LT-STM
working under UHV conditions (see Supplemental Material and reference 11 therein). In our experimental set-up, the plasma
generator is not in the close vicinity of the sample, so that only N
radicals (N*) with thermal energy (and no accelerating voltage)
interact with graphene. Thus, only the topmost layer is affected by
the plasma treatment. This post-synthesis doping method has the advantage over direct growth of modified layers that a well-defined domain could be doped, other part of the sample being kept pristine. 

\begin{figure}[h]
\begin{center}
\includegraphics[width=1\columnwidth]{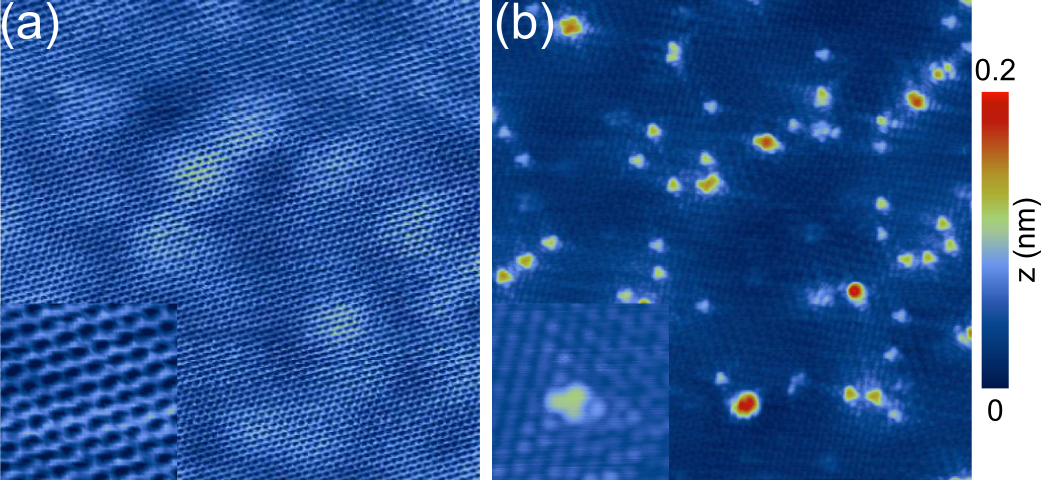}
\caption{(Color online) Comparison of STM images (15$\times$15 nm$^2$) of graphene before
  (a) and after (b) N* exposure. Inset of (a): honeycomb lattice of pristine graphene. Inset of (b): 2$\times$2 nm$^2$ image of a nitrogen dopant. Tunneling parameters: (a) V\subscript{s}=-0.3~V, I=15~nA; (b) V\subscript{s}=-0.5 V, I=500 pA}
\label{grandeEchelle}
\end{center}
\end{figure}

Fig.~\ref{grandeEchelle} presents the graphene surface
before (a) and after (b) N* exposure. A defect-free honeycomb lattice with an
interatomic distance of 1.4~\AA~is observed for the pristine
graphene. After N treatment, many localized features appear as bright dots on the images,
demonstrating the effect of the exposure to nitrogen
radicals. Moreover, Auger Electron Spectroscopy (AES) confirms that the defects are
related to nitrogen doping sites (no other chemical species
(e.g. oxygen) are detected in AES) while Raman spectroscopy confirms
the presence of defects sites after surface treatment (see
Supplemental Material and reference 10 therein). By visual inspection of the STM images, a defect
concentration ranging from 0.5\% to 1.1\% is deduced, depending on the exposure
time (0.6\% for the figures of the present paper).

A closer look at Fig.~\ref{grandeEchelle} (b) reveals that different doping configurations are present. Approximately 75\% of them
display a triangular shape consisting in a bright spot (approximately
0.4 nm wide) with a three-fold symmetry (inset of Fig.~\ref{grandeEchelle} (b)). These images are very similar to the one presented in ref~\onlinecite{zhao} and are assigned to substitutional N atoms. No
preferential orientation of the trigonal pattern is observed and,
consequently, both sublattices of the graphene are affected by the N
treatment. The graphene honeycomb pattern remains also unaltered outside of the vicinity of the defect, demonstrating a very local perturbation of the
graphene layer. Other typical high resolution images of this defect
are shown in Fig.~\ref{serie2} (a-e) for biases ranging from
V\subscript{s}=-0.4 to +0.5~$\unit{V}$ (see Supplemental Material for a more
complete overview). We also systematically observe that the
corrugation at the doping site is more pronounced at positive biases, as
compared to images recorded at negative biases. At this stage, the
doping of graphene with N atoms can then be associated to spatially
localized electronic states in the conduction band.

\begin{figure}[!h]
\begin{center}
\includegraphics[width=0.95\columnwidth]{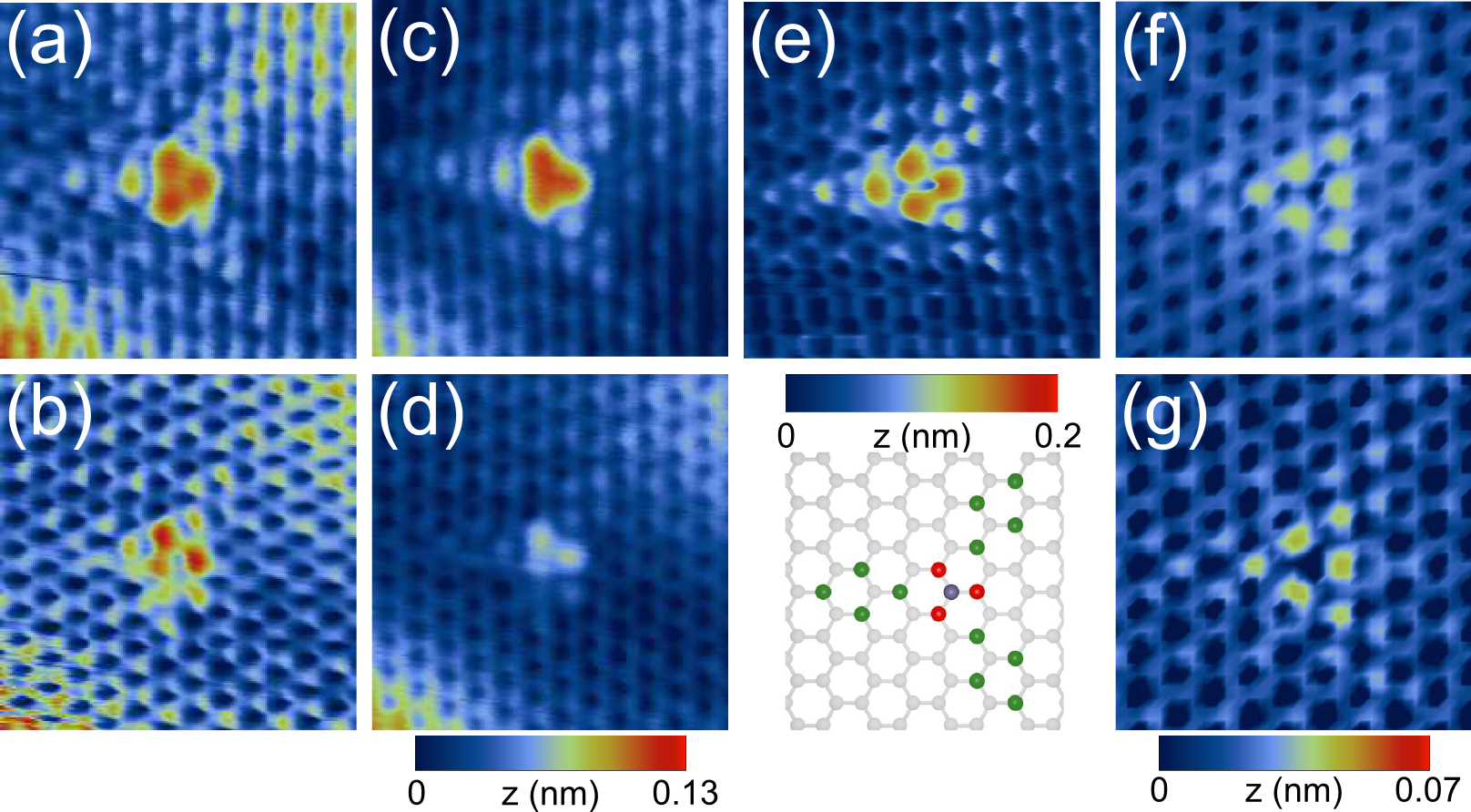}
\caption{(Color online) Topographic images (2.5$\times$2.5 nm$^2$) of a substitutional nitrogen atom at V\subscript{s}=+0.2 V, I=200 pA (a), V\subscript{s}=-0.2 V, I=100 pA (b), V\subscript{s}=+0.5 V, I=700 pA (c), V\subscript{s}=-0.4 V, I=100 pA (d), V\subscript{s}=+0.35 V, I=800 pA (e). In the schematic view, the central small dot (blue online) corresponds to the nitrogen atom, darker and lighter dots (red and green online) correspond to the carbon atomic sites around which the density of states is higher with decreasing values from dark to light (red to green online) as suggested by the experimental images. (f)-(g) Simulations for N substitution in a 10x10 graphene supercell at V\subscript{s}=+0.5 V (f) and V\subscript{s}=-0.5 V (g)}
\label{serie2}
\end{center}
\end{figure}

{\it Ab initio} calculations were performed to obtain the local electronic structure and the STM images using a Tersoff-Hamann approach. Calculations have been performed on a (9$\times$9) supercell (0.6\% of N atoms) and a (10$\times$10) supercell (0.5\% of N atoms) with localized
basis as implemented in the SIESTA package~\cite{sanchezportal} (see
more details in the Supplemental Material and references 12-19 therein). The simulated patterns
(Fig.~\ref{serie2}(f-g); see also refs~\onlinecite{zheng,zhao}) present a dark (low)
spot above the doping atom, surrounded by bright dots,
corresponding to the adjacent C atoms. The central dark spot has
been explained by charge transfer from the N atom to the
neighboring C atoms which results in a smaller spatial extension of
the electronic states associated with the N atom in the direction
perpendicular to the layer compared to the one associated with the C
atoms forming the C-N bonds~\cite{zheng}. This central `hole' is not
observed in most STM images (Fig.~\ref{serie2} and ref~\onlinecite{zhao}) but both the weak dependence of the pattern shape with
the bias voltage and the more intense corrugation for positive biases
are reproduced by simulations. Generally speaking, the limit of our
computation procedure, besides the intrinsic approximations of DFT,
are the simplified tip (a `metallic' `s' shape atomic orbital) and,
more important in the present study, the small distance between the
tip and the atomic layer. For numerical reasons, this distance is
smaller than 3-4~{\AA} in the simulations i.e. close to point contact
and underestimated with respect to the experimental distance.

Interestingly, the hole at the center of the defect (the N atom in
chemical substitution) appears in some of the experimental images, like Fig.~\ref{serie2}(b), i.e. at low negative bias
(see also Supplemental Material). By varying more extensively the
experimental conditions, the same pattern has been observed for
V\subscript{s}=+0.35~V and high current (Fig.~\ref{serie2}(e)). In all these
cases (low V\subscript{s} or high current), the tip is close to the
surface and a better agreement with the simulations is expected,
and is indeed observed. The tip is also closer to the surface for negative bias than for positive bias because
of the localization of the energy state associated with the defect in
the conduction band (see below). This explains why a plain triangular
pattern is observed for V\subscript{s}=+0.5~V whereas a hollow pattern is
observed at V\subscript{s}=-0.2~V.

\begin{figure}[!h]
\begin{center}
\includegraphics[width=1\columnwidth]{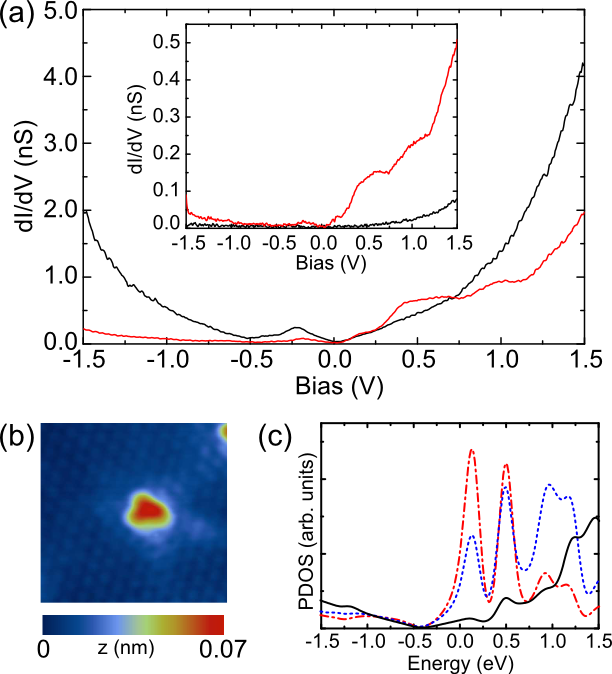}
\caption{(Color online) Comparison of scanning tunneling spectra between graphene
  (black curve) and the simple substitution (gray curve, red
  online). (a): spectra taken with the feedback loop active when
  moving from one spot to another. Inset: spectra taken with the
  feedback loop off when moving from one spot to another.
  (b): Topographic image of the defect on which the
  spectra on (a) have been taken. Tunneling parameters:
  V\subscript{s}=+1.0 V, I=500 pA. (c) Simulated partial-DOS for
  graphene 9$\times$9 supercell (162 atoms) including a single N substitution. The PDOS far away from the nitrogen atom (C\subscript{bulk}), on    the neighboring C atoms (C\subscript{1}) and on the nitrogen atom are represented by the solid, dashed and dot-dashed
  (black, blue and red online) curves, respectively.}
\label{spectro}
\end{center}
\end{figure}

Further insight regarding the local electronic structure of the dopant
can be gained through STS spectra (Fig.~\ref{spectro} (a) with the
corresponding image on fig.~\ref{spectro} (b)). The spectra display
two minima: one at the Fermi energy (0~V), associated with phonon-mediated inelastic
channels~\cite{zhao,zhang}, and one corresponding to the Dirac energy
at -0.5~V. The relationship between the charge carrier concentration
$n$, the Dirac energy $E_D$ and the Fermi Velocity ($v_F$)
$n=\frac{E_D^2}{\pi(\hbar v_F)^2}$ leads to $n=18.10^{12}$ electrons
per cm$²$ for $v_F=10^6 \, \unit{m/s}$. Assuming 0.6\% of nitrogen atoms, the
charge transfer can be estimated to 0.8 electron per dopant atom. From
a theoretical viewpoint, our simulations for a 0.6\% doping (Fig.~\ref{spectro} (c)) exhibit $E_D$ at -0.42 eV and then a smaller charge
transfer (0.55 electron per N atoms). These values are larger than
those reported by Zhao \textit{et al.}~\cite{zhao} and the discrepancy
can be explained by a number of potential sources of uncertainties: the
accuracy in the determination of the nitrogen concentration as well as the position of the Dirac point, the
presence of several types of doping sites and the uncertainty on the
value of the Fermi velocity~\cite{hicks,miller} on the experimental
side or the absence of quasiparticle corrections\footnoteremember{quasi}{Quasiparticle corrections have been demonstrated to lead to a 15\% correction of the Fermi velocity of graphene~\cite{trevisanutto}. Even if the GW calculations have not been performed here due to the large size of the supercell, the correction is expected to be small.} and the regular spacing between the N sites in the supercell technique
for the DFT calculations.

The comparison between the $dI/dV$ spectra at the nitrogen doping
sites and far from it (Fig.~\ref{spectro} (a)), shows that a
broad peak centered around +0.5~eV appears in the vicinity of
the nitrogen atom ($dI/dV$ spectra give a quantity proportional to the
Local Density of State (LDOS)). To the best of our knowledge, this is
the first experimental determination of the energy level of the
localized state in N-doped graphene. The simulated Partial Density of
States (PDOS) located on the N atom, on a C atom close to and far from the nitrogen (C\subscript{1} and C\subscript{bulk} atoms respectivily) are displayed on Fig.~\ref{spectro}(c). The states related to the graphitic
nitrogen are clearly obtained in the conduction bands but display a
double peak structure at 0.15 eV and 0.50 eV. The nitrogen-nitrogen interactions in the periodical structure used in the calculation could lead to a splitting of the donor state that would not be present for randomly distributed defects~\cite{lambin}. The absence of quasiparticle corrections\footnoterecall{quasi} or the reduced tunneling current for low bias~\cite{zhang} may also explain why the low energy states are not observed. Besides, experimental
STS spectra obtained for a tip above the center of the defect probably
also probe the PDOS of the C\subscript{1} atoms. Like the N PDOS, the C\subscript{1} PDOS presents a double peak at
0.15 eV and 0.50 eV but with larger amplitude of the second
one. Moreover a state localized at the C\subscript{1} atoms appears around 1 eV
and can be related to a second feature in the experimental STS.

Coming back to the experimental STS spectra, at negative
bias, the $dI/dV$ signal is found to be lower above the nitrogen atom than above
the graphene as shown in Fig.~\ref{spectro}(a). At positive bias both
STS spectra have comparable intensities. To reconcile those measurements
with the higher corrugation above the dopant site (Figs~\ref{grandeEchelle} and \ref{serie2}), we have to keep in mind that
spectra are measured with initial conditions corresponding to
the setpoint used for the STM image. As a consequence, the tip is located
at a larger distance from the atomic plane when a spectrum is
measured above the dopant site, compared with a STS measurement on the
graphene layer and the two spectra cannot be quantitatively compared. As
it is expected that the nitrogen atom lies in the plane of the
graphene sheet, as demonstrated by DFT simulations performed on single~\cite{zheng} and bilayer N-doped graphene~\cite{guillaume}, the
observed corrugation is a purely electronic effect. In such cases,
the intensity of $dI/dV$ spectra are artificially reduced above the N atom due to the higher tip position. In order to
overcome this difficulty of interpretation, we have measured some
spectra while scanning with the STM feedback loop off, i.e. at a constant tip height. In these conditions, STS spectra exhibit a very different behavior (inset in Fig.~\ref{spectro} (a)). For negative bias, the spectrum recorded above the
dopant is slightly larger than above the graphene layer, in
agreement with what is observed on the images. For positive bias, the difference is even more spectacular
with a LDOS measured at the doping site up to five times more intense
than the one of graphene, strongly confirming that the doping states
lie in the conduction band.

\begin{figure}[!h]
\begin{center}
\includegraphics[width=0.95\columnwidth]{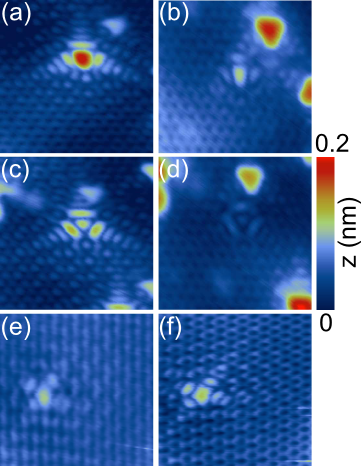}
\caption{(Color online) STM images of complex doping sites (3$\times$3 nm$^2$). (a), (c) and (e) correspond to three different defects measured at a bias voltage of -0.5 V. (b), (d) and (f) correspond respectively to the same defects measured at +0.5 V.}
\label{defect2}
\end{center}
\end{figure}

Besides the substitutional doping, other types of defects have been
frequently observed, examples of which are displayed in Fig.~\ref{defect2}. Some
defects appear higher at negative bias (Fig.~\ref{defect2} (a) and (c))
than at positive bias (Fig.~\ref{defect2} (b) and (d)) contrary to what was observed for substitution. Interestingly, a simple
substitutional N atom is also observed on top of Fig.~\ref{defect2} (a)
to (d) close to the complex defects. These configurations allow a direct comparison
of the bias dependence of the simple substitutional N (higher at
positive bias) and of the other defects (higher at negative bias).
These new dopant sites can then be associated with hole-doping and with a localized state lying in the valence band~\cite{ewels,zheng}. Furthermore, the experimental signature (a
triangular symmetry with extended oscillations of the wavefunction) is
consistent with the behavior predicted for a configuration involving
three N atoms forming a pyridinic configuration~\cite{zheng} or with a single vacancy~\cite{amara}.
Other defects, like the one reported in Fig.~\ref{defect2} (e) and (f) present
a pattern with a rectangular symmetry and exhibit a n-type
characteristic (higher corrugation at positive bias). However, further
investigations based on STS analysis are necessary to reach a definitive conclusion on the atomic configurations of these complex doping sites.

In summary, the N doping of graphene was achieved by exposing epitaxial samples to an atomic nitrogen flux. The most common ($\sim$75\%) doping configuration is found to be substitutional (graphitic) N atoms. Local STS Spectra have provided the first experimental evidence of a localized donor state, specific signature of N substitution. STM topographical images revealed a depletion above the N sites characteristic of a charge transfer between the N atom and the adjacent neighboring C atoms. Finally, other N doping sites with more complex atomic structure than a simple substitution were observed, exhibiting an acceptor-like character. The presented post-synthesis treatment opens the door to local tuning of electronic properties of graphene which is a prerequisite for the development of carbon-based electronics. 

\section{Acknowledgments}

F. J. thanks Jessica Campos Delgado and Beno\^{i}t Hackens for the Raman
measurements and Jacques Ghijsen for proofreading of the manuscript. We acknowledge fruitful discussions with Philippe
Lambin, Fran\c cois Ducastelle and Hakim Amara. This research used
resources of the Interuniversity Scientific Computing Facility located
at the University of Namur, Belgium, which is supported by the
F.R.S.-FNRS under convention N$^{\circ}$ 2.4617.07. V. R. thanks the
Institut Universitaire de France for support. This work is directly connected to the Belgian Program on Interuniversity Attraction Poles (PAI6) on ``Quantum Effects in Clusters and Nanowires'' and to the ARC on ``Graphene StressTronics'' (Convention N$^{\circ}$ 11/16-037) sponsored by the Communaut\'{e} Fran\c caise de Belgique, and to the European Commission through the ETSF e-I3 project (Grant No. 211956).

\def\bibfont{\footnotesize}
\bibliography{../../../Biblio/graphene}

\begin{thebibliography}{31}
\expandafter\ifx\csname natexlab\endcsname\relax\def\natexlab#1{#1}\fi
\expandafter\ifx\csname bibnamefont\endcsname\relax
  \def\bibnamefont#1{#1}\fi
\expandafter\ifx\csname bibfnamefont\endcsname\relax
  \def\bibfnamefont#1{#1}\fi
\expandafter\ifx\csname citenamefont\endcsname\relax
  \def\citenamefont#1{#1}\fi
\expandafter\ifx\csname url\endcsname\relax
  \def\url#1{\texttt{#1}}\fi
\expandafter\ifx\csname urlprefix\endcsname\relax\def\urlprefix{URL }\fi
\providecommand{\bibinfo}[2]{#2}
\providecommand{\eprint}[2][]{\url{#2}}

\bibitem[{\citenamefont{Castro~Neto et~al.}(2009)\citenamefont{Castro~Neto,
  Guinea, Peres, Novoselov, and Geim}}]{neto}
\bibinfo{author}{\bibfnamefont{A.~H.} \bibnamefont{Castro~Neto}},
  \bibinfo{author}{\bibfnamefont{F.}~\bibnamefont{Guinea}},
  \bibinfo{author}{\bibfnamefont{N.~M.~R.} \bibnamefont{Peres}},
  \bibinfo{author}{\bibfnamefont{K.~S.} \bibnamefont{Novoselov}},
  \bibnamefont{and} \bibinfo{author}{\bibfnamefont{A.~K.} \bibnamefont{Geim}},
  \bibinfo{journal}{Rev. Mod. Phys.} \textbf{\bibinfo{volume}{81}},
  \bibinfo{pages}{109} (\bibinfo{year}{2009}).

\bibitem[{\citenamefont{Liu et~al.}(2011)\citenamefont{Liu, Liu, and
  Zhu}}]{liu}
\bibinfo{author}{\bibfnamefont{H.}~\bibnamefont{Liu}},
  \bibinfo{author}{\bibfnamefont{Y.}~\bibnamefont{Liu}}, \bibnamefont{and}
  \bibinfo{author}{\bibfnamefont{D.}~\bibnamefont{Zhu}},
  \bibinfo{journal}{Journal of Materials Chemistry}
  \textbf{\bibinfo{volume}{21}}, \bibinfo{pages}{3335} (\bibinfo{year}{2011}).

\bibitem[{\citenamefont{Fuhrer et~al.}(2010)\citenamefont{Fuhrer, Lau, and
  MacDonald}}]{fuhrer}
\bibinfo{author}{\bibfnamefont{M.~S.} \bibnamefont{Fuhrer}},
  \bibinfo{author}{\bibfnamefont{C.~N.} \bibnamefont{Lau}}, \bibnamefont{and}
  \bibinfo{author}{\bibfnamefont{A.~H.} \bibnamefont{MacDonald}},
  \bibinfo{journal}{MRS Bulletin} \textbf{\bibinfo{volume}{35}},
  \bibinfo{pages}{289} (\bibinfo{year}{2010}).

\bibitem[{\citenamefont{Ewels and Glerup}(2005)}]{ewels}
\bibinfo{author}{\bibfnamefont{C.~P.} \bibnamefont{Ewels}} \bibnamefont{and}
  \bibinfo{author}{\bibfnamefont{M.}~\bibnamefont{Glerup}},
  \bibinfo{journal}{J. Nanosci. Nanotechnol.} \textbf{\bibinfo{volume}{5}},
  \bibinfo{pages}{1345} (\bibinfo{year}{2005}).

\bibitem[{\citenamefont{Lherbier et~al.}(2008)\citenamefont{Lherbier, Blase,
  Niquet, Triozon, and Roche}}]{lherbier}
\bibinfo{author}{\bibfnamefont{A.}~\bibnamefont{Lherbier}},
  \bibinfo{author}{\bibfnamefont{X.}~\bibnamefont{Blase}},
  \bibinfo{author}{\bibfnamefont{Y.-M.} \bibnamefont{Niquet}},
  \bibinfo{author}{\bibfnamefont{F.}~\bibnamefont{Triozon}}, \bibnamefont{and}
  \bibinfo{author}{\bibfnamefont{S.}~\bibnamefont{Roche}},
  \bibinfo{journal}{Phys. Rev. Lett.} \textbf{\bibinfo{volume}{101}},
  \bibinfo{pages}{036808} (\bibinfo{year}{2008}).

\bibitem[{\citenamefont{Zheng et~al.}(2010)\citenamefont{Zheng, Hermet, and
  Henrard}}]{zheng}
\bibinfo{author}{\bibfnamefont{B.}~\bibnamefont{Zheng}},
  \bibinfo{author}{\bibfnamefont{P.}~\bibnamefont{Hermet}}, \bibnamefont{and}
  \bibinfo{author}{\bibfnamefont{L.}~\bibnamefont{Henrard}},
  \bibinfo{journal}{ACS Nano} \textbf{\bibinfo{volume}{7}},
  \bibinfo{pages}{4165} (\bibinfo{year}{2010}).

\bibitem[{\citenamefont{Wang et~al.}(2010)\citenamefont{Wang, Shao, Watson, Li,
  and Lin}}]{wang}
\bibinfo{author}{\bibfnamefont{Y.}~\bibnamefont{Wang}},
  \bibinfo{author}{\bibfnamefont{Y.}~\bibnamefont{Shao}},
  \bibinfo{author}{\bibfnamefont{D.}~\bibnamefont{Watson}},
  \bibinfo{author}{\bibfnamefont{J.}~\bibnamefont{Li}}, \bibnamefont{and}
  \bibinfo{author}{\bibfnamefont{Y.}~\bibnamefont{Lin}}, \bibinfo{journal}{ACS
  Nano} \textbf{\bibinfo{volume}{4}}, \bibinfo{pages}{1790}
  (\bibinfo{year}{2010}).

\bibitem[{\citenamefont{Soin et~al.}(2011)\citenamefont{Soin, Roy, Roy, Hazra,
  Misra, Lim, Hetherington, and McLaughlin}}]{soin}
\bibinfo{author}{\bibfnamefont{N.}~\bibnamefont{Soin}},
  \bibinfo{author}{\bibfnamefont{S.~S.} \bibnamefont{Roy}},
  \bibinfo{author}{\bibfnamefont{S.}~\bibnamefont{Roy}},
  \bibinfo{author}{\bibfnamefont{K.~S.} \bibnamefont{Hazra}},
  \bibinfo{author}{\bibfnamefont{D.~S.} \bibnamefont{Misra}},
  \bibinfo{author}{\bibfnamefont{T.~H.} \bibnamefont{Lim}},
  \bibinfo{author}{\bibfnamefont{C.~J.} \bibnamefont{Hetherington}},
  \bibnamefont{and} \bibinfo{author}{\bibfnamefont{J.~A.}
  \bibnamefont{McLaughlin}}, \bibinfo{journal}{J. Phys. Chem. C}
  \textbf{\bibinfo{volume}{115}}, \bibinfo{pages}{5366} (\bibinfo{year}{2011}).

\bibitem[{\citenamefont{Reddy et~al.}(2010)\citenamefont{Reddy, Srivastava,
  Gowda, Gullapalli, Dubey, and Ajayan}}]{reddy}
\bibinfo{author}{\bibfnamefont{A.~L.~M.} \bibnamefont{Reddy}},
  \bibinfo{author}{\bibfnamefont{A.}~\bibnamefont{Srivastava}},
  \bibinfo{author}{\bibfnamefont{S.~R.} \bibnamefont{Gowda}},
  \bibinfo{author}{\bibfnamefont{H.}~\bibnamefont{Gullapalli}},
  \bibinfo{author}{\bibfnamefont{M.}~\bibnamefont{Dubey}}, \bibnamefont{and}
  \bibinfo{author}{\bibfnamefont{P.~M.} \bibnamefont{Ajayan}},
  \bibinfo{journal}{ACS nano} \textbf{\bibinfo{volume}{4}},
  \bibinfo{pages}{6337} (\bibinfo{year}{2010}).

\bibitem[{\citenamefont{Cho et~al.}(2011)\citenamefont{Cho, Kim, Im, Myung,
  Jung, Lee, Park, Park, Cho, and Kang}}]{cho}
\bibinfo{author}{\bibfnamefont{Y.~J.} \bibnamefont{Cho}},
  \bibinfo{author}{\bibfnamefont{H.~S.} \bibnamefont{Kim}},
  \bibinfo{author}{\bibfnamefont{H.}~\bibnamefont{Im}},
  \bibinfo{author}{\bibfnamefont{Y.}~\bibnamefont{Myung}},
  \bibinfo{author}{\bibfnamefont{G.~B.} \bibnamefont{Jung}},
  \bibinfo{author}{\bibfnamefont{C.}~\bibnamefont{Lee}},
  \bibinfo{author}{\bibfnamefont{J.}~\bibnamefont{Park}},
  \bibinfo{author}{\bibfnamefont{M.-H.} \bibnamefont{Park}},
  \bibinfo{author}{\bibfnamefont{J.}~\bibnamefont{Cho}}, \bibnamefont{and}
  \bibinfo{author}{\bibfnamefont{H.~S.} \bibnamefont{Kang}},
  \bibinfo{journal}{The Journal of Physical Chemistry C}
  \textbf{\bibinfo{volume}{115}}, \bibinfo{pages}{9451} (\bibinfo{year}{2011}).

\bibitem[{\citenamefont{Kasry et~al.}(2010)\citenamefont{Kasry, Kuroda,
  Martyna, Tulevski, and a.~Bol}}]{kasry}
\bibinfo{author}{\bibfnamefont{A.}~\bibnamefont{Kasry}},
  \bibinfo{author}{\bibfnamefont{M.~A.} \bibnamefont{Kuroda}},
  \bibinfo{author}{\bibfnamefont{G.}~\bibnamefont{Martyna}},
  \bibinfo{author}{\bibfnamefont{G.}~\bibnamefont{Tulevski}}, \bibnamefont{and}
  \bibinfo{author}{\bibfnamefont{A.}~\bibnamefont{a.~Bol}},
  \bibinfo{journal}{ACS Nano} \textbf{\bibinfo{volume}{4 (7)}},
  \bibinfo{pages}{3839} (\bibinfo{year}{2010}).

\bibitem[{\citenamefont{Wei et~al.}(2009)\citenamefont{Wei, Liu, Wang, Zhang,
  Huang, and Yu}}]{wei}
\bibinfo{author}{\bibfnamefont{D.}~\bibnamefont{Wei}},
  \bibinfo{author}{\bibfnamefont{Y.}~\bibnamefont{Liu}},
  \bibinfo{author}{\bibfnamefont{Y.}~\bibnamefont{Wang}},
  \bibinfo{author}{\bibfnamefont{H.}~\bibnamefont{Zhang}},
  \bibinfo{author}{\bibfnamefont{L.}~\bibnamefont{Huang}}, \bibnamefont{and}
  \bibinfo{author}{\bibfnamefont{G.}~\bibnamefont{Yu}}, \bibinfo{journal}{Nano
  Lett.} \textbf{\bibinfo{volume}{9}}, \bibinfo{pages}{1752}
  (\bibinfo{year}{2009}).

\bibitem[{\citenamefont{Zhao et~al.}(2011)\citenamefont{Zhao, He, Rim, Schiros,
  Kim, Zhou, Gutiérrez, Chockalingam, Arguello, Pálová et~al.}}]{zhao}
\bibinfo{author}{\bibfnamefont{L.}~\bibnamefont{Zhao}},
  \bibinfo{author}{\bibfnamefont{R.}~\bibnamefont{He}},
  \bibinfo{author}{\bibfnamefont{K.~T.} \bibnamefont{Rim}},
  \bibinfo{author}{\bibfnamefont{T.}~\bibnamefont{Schiros}},
  \bibinfo{author}{\bibfnamefont{K.~S.} \bibnamefont{Kim}},
  \bibinfo{author}{\bibfnamefont{H.}~\bibnamefont{Zhou}},
  \bibinfo{author}{\bibfnamefont{C.}~\bibnamefont{Gutiérrez}},
  \bibinfo{author}{\bibfnamefont{S.~P.} \bibnamefont{Chockalingam}},
  \bibinfo{author}{\bibfnamefont{C.~J.} \bibnamefont{Arguello}},
  \bibinfo{author}{\bibfnamefont{L.}~\bibnamefont{Pálová}},
  \bibnamefont{et~al.}, \bibinfo{journal}{Science}
  \textbf{\bibinfo{volume}{333}}, \bibinfo{pages}{999} (\bibinfo{year}{2011}).

\bibitem[{\citenamefont{Panchakarla et~al.}(2009)\citenamefont{Panchakarla,
  Subrahmanyam, Saha, Govindaraj, Krishnamurthy, Waghmare, and
  Rao}}]{panchakarla}
\bibinfo{author}{\bibfnamefont{L.~S.} \bibnamefont{Panchakarla}},
  \bibinfo{author}{\bibfnamefont{K.~S.} \bibnamefont{Subrahmanyam}},
  \bibinfo{author}{\bibfnamefont{S.~K.} \bibnamefont{Saha}},
  \bibinfo{author}{\bibfnamefont{A.}~\bibnamefont{Govindaraj}},
  \bibinfo{author}{\bibfnamefont{H.~R.} \bibnamefont{Krishnamurthy}},
  \bibinfo{author}{\bibfnamefont{U.~V.} \bibnamefont{Waghmare}},
  \bibnamefont{and} \bibinfo{author}{\bibfnamefont{C.~N.~R.}
  \bibnamefont{Rao}}, \bibinfo{journal}{Advanced Materials}
  \textbf{\bibinfo{volume}{26}}, \bibinfo{pages}{4726} (\bibinfo{year}{2009}).

\bibitem[{\citenamefont{Meyer et~al.}(2011)\citenamefont{Meyer, Kurasch, Park,
  Skakalova, Künzel, Groß, Chuvilin, Algara-Siller, Roth, Iwasaki
  et~al.}}]{meyer}
\bibinfo{author}{\bibfnamefont{J.~C.} \bibnamefont{Meyer}},
  \bibinfo{author}{\bibfnamefont{S.}~\bibnamefont{Kurasch}},
  \bibinfo{author}{\bibfnamefont{H.~J.} \bibnamefont{Park}},
  \bibinfo{author}{\bibfnamefont{V.}~\bibnamefont{Skakalova}},
  \bibinfo{author}{\bibfnamefont{D.}~\bibnamefont{Künzel}},
  \bibinfo{author}{\bibfnamefont{A.}~\bibnamefont{Groß}},
  \bibinfo{author}{\bibfnamefont{A.}~\bibnamefont{Chuvilin}},
  \bibinfo{author}{\bibfnamefont{G.}~\bibnamefont{Algara-Siller}},
  \bibinfo{author}{\bibfnamefont{S.}~\bibnamefont{Roth}},
  \bibinfo{author}{\bibfnamefont{T.}~\bibnamefont{Iwasaki}},
  \bibnamefont{et~al.}, \bibinfo{journal}{Nat Mater}
  \textbf{\bibinfo{volume}{10}}, \bibinfo{pages}{209} (\bibinfo{year}{2011}).

\bibitem[{\citenamefont{Guo et~al.}(2010)\citenamefont{Guo, Liu, Chen, Zhu,
  Fang, and Gong}}]{guo}
\bibinfo{author}{\bibfnamefont{B.}~\bibnamefont{Guo}},
  \bibinfo{author}{\bibfnamefont{Q.}~\bibnamefont{Liu}},
  \bibinfo{author}{\bibfnamefont{E.}~\bibnamefont{Chen}},
  \bibinfo{author}{\bibfnamefont{H.}~\bibnamefont{Zhu}},
  \bibinfo{author}{\bibfnamefont{L.}~\bibnamefont{Fang}}, \bibnamefont{and}
  \bibinfo{author}{\bibfnamefont{J.~R.} \bibnamefont{Gong}},
  \bibinfo{journal}{Nano Lett.} \textbf{\bibinfo{volume}{10}},
  \bibinfo{pages}{4975} (\bibinfo{year}{2010}).

\bibitem[{\citenamefont{Lin et~al.}(2010)\citenamefont{Lin, Lin, and
  Chiu}}]{lin}
\bibinfo{author}{\bibfnamefont{Y.-C.} \bibnamefont{Lin}},
  \bibinfo{author}{\bibfnamefont{C.-Y.} \bibnamefont{Lin}}, \bibnamefont{and}
  \bibinfo{author}{\bibfnamefont{P.-W.} \bibnamefont{Chiu}},
  \bibinfo{journal}{Appl. phys. Lett.} \textbf{\bibinfo{volume}{96}},
  \bibinfo{pages}{133110} (\bibinfo{year}{2010}).

\bibitem[{\citenamefont{Jeong et~al.}(2011)\citenamefont{Jeong, Lee, Shin,
  Choi, Shin, Kang, and Choi}}]{jeong}
\bibinfo{author}{\bibfnamefont{H.~M.} \bibnamefont{Jeong}},
  \bibinfo{author}{\bibfnamefont{J.~W.} \bibnamefont{Lee}},
  \bibinfo{author}{\bibfnamefont{W.~H.} \bibnamefont{Shin}},
  \bibinfo{author}{\bibfnamefont{Y.~J.} \bibnamefont{Choi}},
  \bibinfo{author}{\bibfnamefont{H.~J.} \bibnamefont{Shin}},
  \bibinfo{author}{\bibfnamefont{J.~K.} \bibnamefont{Kang}}, \bibnamefont{and}
  \bibinfo{author}{\bibfnamefont{J.~W.} \bibnamefont{Choi}},
  \bibinfo{journal}{Nano Lett.} \textbf{\bibinfo{volume}{11}},
  \bibinfo{pages}{2472} (\bibinfo{year}{2011}).

\bibitem[{\citenamefont{Forbeaux et~al.}(1999)\citenamefont{Forbeaux, Themlin,
  and Debever}}]{forbeaux}
\bibinfo{author}{\bibfnamefont{I.}~\bibnamefont{Forbeaux}},
  \bibinfo{author}{\bibfnamefont{J.-M.} \bibnamefont{Themlin}},
  \bibnamefont{and} \bibinfo{author}{\bibfnamefont{J.-M.}
  \bibnamefont{Debever}}, \bibinfo{journal}{Surf. Science}
  \textbf{\bibinfo{volume}{442}}, \bibinfo{pages}{9} (\bibinfo{year}{1999}).

\bibitem[{\citenamefont{de~Heer et~al.}(2007)\citenamefont{de~Heer, Berger,
  Wua, First, Conrada, Li, Li, Sprinklea, Hassa, Sadowski et~al.}}]{deheer}
\bibinfo{author}{\bibfnamefont{W.~A.} \bibnamefont{de~Heer}},
  \bibinfo{author}{\bibfnamefont{C.}~\bibnamefont{Berger}},
  \bibinfo{author}{\bibfnamefont{X.}~\bibnamefont{Wua}},
  \bibinfo{author}{\bibfnamefont{P.~N.} \bibnamefont{First}},
  \bibinfo{author}{\bibfnamefont{E.~H.} \bibnamefont{Conrada}},
  \bibinfo{author}{\bibfnamefont{X.}~\bibnamefont{Li}},
  \bibinfo{author}{\bibfnamefont{T.}~\bibnamefont{Li}},
  \bibinfo{author}{\bibfnamefont{M.}~\bibnamefont{Sprinklea}},
  \bibinfo{author}{\bibfnamefont{J.}~\bibnamefont{Hassa}},
  \bibinfo{author}{\bibfnamefont{M.~L.} \bibnamefont{Sadowski}},
  \bibnamefont{et~al.}, \bibinfo{journal}{Solid State Commun.}
  \textbf{\bibinfo{volume}{143}}, \bibinfo{pages}{92} (\bibinfo{year}{2007}).

\bibitem[{\citenamefont{Varchon et~al.}(2008)\citenamefont{Varchon, Mallet,
  Magaud, and Veuillen}}]{varchon}
\bibinfo{author}{\bibfnamefont{F.}~\bibnamefont{Varchon}},
  \bibinfo{author}{\bibfnamefont{P.}~\bibnamefont{Mallet}},
  \bibinfo{author}{\bibfnamefont{L.}~\bibnamefont{Magaud}}, \bibnamefont{and}
  \bibinfo{author}{\bibfnamefont{J.-Y.} \bibnamefont{Veuillen}},
  \bibinfo{journal}{Phys. Rev. B} \textbf{\bibinfo{volume}{77}},
  \bibinfo{pages}{165415} (\bibinfo{year}{2008}).

\bibitem[{\citenamefont{Hiebel et~al.}(2008)\citenamefont{Hiebel, Mallet,
  Varchon, Magaud, and Veuillen}}]{hiebel}
\bibinfo{author}{\bibfnamefont{F.}~\bibnamefont{Hiebel}},
  \bibinfo{author}{\bibfnamefont{P.}~\bibnamefont{Mallet}},
  \bibinfo{author}{\bibfnamefont{F.}~\bibnamefont{Varchon}},
  \bibinfo{author}{\bibfnamefont{L.}~\bibnamefont{Magaud}}, \bibnamefont{and}
  \bibinfo{author}{\bibfnamefont{J.-Y.} \bibnamefont{Veuillen}},
  \bibinfo{journal}{Phys. Rev. B} \textbf{\bibinfo{volume}{78}},
  \bibinfo{pages}{153412} (\bibinfo{year}{2008}).

\bibitem[{\citenamefont{Starke and Riedl}(2009)}]{starke}
\bibinfo{author}{\bibfnamefont{U.}~\bibnamefont{Starke}} \bibnamefont{and}
  \bibinfo{author}{\bibfnamefont{C.}~\bibnamefont{Riedl}}, \bibinfo{journal}{J.
  Phys.: Condens. Matter} \textbf{\bibinfo{volume}{21}},
  \bibinfo{pages}{134016} (\bibinfo{year}{2009}).

\bibitem[{\citenamefont{S\'anchez-Portal
  et~al.}(1997)\citenamefont{S\'anchez-Portal, Ordej\'on, Artacho, and
  Soler}}]{sanchezportal}
\bibinfo{author}{\bibfnamefont{D.}~\bibnamefont{S\'anchez-Portal}},
  \bibinfo{author}{\bibfnamefont{P.}~\bibnamefont{Ordej\'on}},
  \bibinfo{author}{\bibfnamefont{E.}~\bibnamefont{Artacho}}, \bibnamefont{and}
  \bibinfo{author}{\bibfnamefont{J.~M.} \bibnamefont{Soler}},
  \bibinfo{journal}{Int. J. Quantum Chem.} \textbf{\bibinfo{volume}{65}},
  \bibinfo{pages}{453} (\bibinfo{year}{1997}).

\bibitem[{\citenamefont{Zhang et~al.}(2008)\citenamefont{Zhang, Brar, Wang,
  Girit, Yayon, Panlasigui, Zettl, and Crommie}}]{zhang}
\bibinfo{author}{\bibfnamefont{Y.}~\bibnamefont{Zhang}},
  \bibinfo{author}{\bibfnamefont{V.~W.} \bibnamefont{Brar}},
  \bibinfo{author}{\bibfnamefont{F.}~\bibnamefont{Wang}},
  \bibinfo{author}{\bibfnamefont{C.}~\bibnamefont{Girit}},
  \bibinfo{author}{\bibfnamefont{Y.}~\bibnamefont{Yayon}},
  \bibinfo{author}{\bibfnamefont{M.}~\bibnamefont{Panlasigui}},
  \bibinfo{author}{\bibfnamefont{A.}~\bibnamefont{Zettl}}, \bibnamefont{and}
  \bibinfo{author}{\bibfnamefont{M.~F.} \bibnamefont{Crommie}},
  \bibinfo{journal}{Nat Phys} \textbf{\bibinfo{volume}{4}},
  \bibinfo{pages}{627} (\bibinfo{year}{2008}).

\bibitem[{\citenamefont{Hicks et~al.}(2011)\citenamefont{Hicks, Sprinkle,
  Shepperd, Wang, Tejeda, Taleb-Ibrahimi, Bertran, Le~F\`evre, de~Heer, Berger
  et~al.}}]{hicks}
\bibinfo{author}{\bibfnamefont{J.}~\bibnamefont{Hicks}},
  \bibinfo{author}{\bibfnamefont{M.}~\bibnamefont{Sprinkle}},
  \bibinfo{author}{\bibfnamefont{K.}~\bibnamefont{Shepperd}},
  \bibinfo{author}{\bibfnamefont{F.}~\bibnamefont{Wang}},
  \bibinfo{author}{\bibfnamefont{A.}~\bibnamefont{Tejeda}},
  \bibinfo{author}{\bibfnamefont{A.}~\bibnamefont{Taleb-Ibrahimi}},
  \bibinfo{author}{\bibfnamefont{F.}~\bibnamefont{Bertran}},
  \bibinfo{author}{\bibfnamefont{P.}~\bibnamefont{Le~F\`evre}},
  \bibinfo{author}{\bibfnamefont{W.~A.} \bibnamefont{de~Heer}},
  \bibinfo{author}{\bibfnamefont{C.}~\bibnamefont{Berger}},
  \bibnamefont{et~al.}, \bibinfo{journal}{Phys. Rev. B}
  \textbf{\bibinfo{volume}{83}}, \bibinfo{pages}{205403}
  (\bibinfo{year}{2011}).

\bibitem[{\citenamefont{Miller et~al.}(2009)\citenamefont{Miller, Kubista,
  Rutter, Ruan, de~Heer, First, and Stroscio}}]{miller}
\bibinfo{author}{\bibfnamefont{D.~L.} \bibnamefont{Miller}},
  \bibinfo{author}{\bibfnamefont{K.~D.} \bibnamefont{Kubista}},
  \bibinfo{author}{\bibfnamefont{G.~M.} \bibnamefont{Rutter}},
  \bibinfo{author}{\bibfnamefont{M.}~\bibnamefont{Ruan}},
  \bibinfo{author}{\bibfnamefont{W.~A.} \bibnamefont{de~Heer}},
  \bibinfo{author}{\bibfnamefont{P.~N.} \bibnamefont{First}}, \bibnamefont{and}
  \bibinfo{author}{\bibfnamefont{J.~A.} \bibnamefont{Stroscio}},
  \bibinfo{journal}{Science} \textbf{\bibinfo{volume}{324}},
  \bibinfo{pages}{924} (\bibinfo{year}{2009}).

\bibitem[{\citenamefont{Lambin et~al.}()\citenamefont{Lambin, Amara,
  Ducastelle, and Henrard}}]{lambin}
\bibinfo{author}{\bibfnamefont{P.}~\bibnamefont{Lambin}},
  \bibinfo{author}{\bibfnamefont{H.}~\bibnamefont{Amara}},
  \bibinfo{author}{\bibfnamefont{F.}~\bibnamefont{Ducastelle}},
  \bibnamefont{and} \bibinfo{author}{\bibfnamefont{L.}~\bibnamefont{Henrard}},
  \bibinfo{note}{in preparation}.

\bibitem[{\citenamefont{Guillaume et~al.}(2012)\citenamefont{Guillaume, Zheng,
  Charlier, and Henrard}}]{guillaume}
\bibinfo{author}{\bibfnamefont{S.-O.} \bibnamefont{Guillaume}},
  \bibinfo{author}{\bibfnamefont{B.}~\bibnamefont{Zheng}},
  \bibinfo{author}{\bibfnamefont{J.-C.} \bibnamefont{Charlier}},
  \bibnamefont{and} \bibinfo{author}{\bibfnamefont{L.}~\bibnamefont{Henrard}},
  \bibinfo{journal}{Phys. Rev. B} \textbf{\bibinfo{volume}{85}},
  \bibinfo{pages}{035444} (\bibinfo{year}{2012}).

\bibitem[{\citenamefont{Amara et~al.}(2007)\citenamefont{Amara, Latil, Meunier,
  Lambin, and Charlier}}]{amara}
\bibinfo{author}{\bibfnamefont{H.}~\bibnamefont{Amara}},
  \bibinfo{author}{\bibfnamefont{S.}~\bibnamefont{Latil}},
  \bibinfo{author}{\bibfnamefont{V.}~\bibnamefont{Meunier}},
  \bibinfo{author}{\bibfnamefont{P.}~\bibnamefont{Lambin}}, \bibnamefont{and}
  \bibinfo{author}{\bibfnamefont{J.-C.} \bibnamefont{Charlier}},
  \bibinfo{journal}{Phys. Rev. B} \textbf{\bibinfo{volume}{76}},
  \bibinfo{pages}{115423} (\bibinfo{year}{2007}).

\bibitem[{\citenamefont{Trevisanutto et~al.}(2008)\citenamefont{Trevisanutto,
  Giorgetti, Reining, Ladisa, and Olevano}}]{trevisanutto}
\bibinfo{author}{\bibfnamefont{P.~E.} \bibnamefont{Trevisanutto}},
  \bibinfo{author}{\bibfnamefont{C.}~\bibnamefont{Giorgetti}},
  \bibinfo{author}{\bibfnamefont{L.}~\bibnamefont{Reining}},
  \bibinfo{author}{\bibfnamefont{M.}~\bibnamefont{Ladisa}}, \bibnamefont{and}
  \bibinfo{author}{\bibfnamefont{V.}~\bibnamefont{Olevano}},
  \bibinfo{journal}{Phys. Rev. Lett.} \textbf{\bibinfo{volume}{101}},
  \bibinfo{pages}{226405} (\bibinfo{year}{2008}).

\end{thebibliography}
\end{document}